\documentclass[showpacs,pre,amsmath,amssymb]{revtex4-1}
\usepackage{graphicx}
\usepackage{bm}
\def\bra{\langle}
\def\ket{\rangle}
\def\bvec#1{\mathbf{#1}}
\def\vb{\bvec{b}}

\def\vf{\bvec{f}}
\def\vx{\bvec{x}}
\def\vk{\bvec{k}}
\def\vp{\bvec{p}}
\def\vq{\bvec{q}}

\def\vu{\bvec{u}}
\def\vy{\bvec{y}}
\def\cD{\mathcal{D}}
\def\cF{\mathcal{F}}

\def\R{\mathbb{R}}

\begin{document}
\title{Renormalization of viscosity in wavelet-based model of turbulence}

\author{M.V.Altaisky}
\affiliation{Space Research Institute RAS, Profsoyuznaya 84/32, Moscow, 117997, Russia}
\email{altaisky@rssi.ru}

\author{M.Hnatich}
\affiliation{P.J.\v{S}af\`{a}rik University in Ko\v{s}ice, Park Angelinum 9, 04154, Ko\v{s}ice, Slovakia and Institute of Experimental Physics SAS, Watsonova 47, 040 01, Ko\v{s}ice, Slovakia}
\affiliation{Joint Institute for Nuclear Research, Joliot-Curie 6, Dubna, 141980, Russia}
\email{hnatic@saske.sk}

\author{N.E.Kaputkina}
\affiliation{National University of Science and Technology 'MISIS',  Leninsky av. 4, Moscow, 119049, Russia}

\email{nataly@misis.ru}

\date{Nov 22, 2017}
\date{\today} 
    
\begin{abstract}
Statistical theory of turbulence in viscid incompressible fluid, 
described by the Navier-Stokes equation driven by random force, is reformulated in terms of scale-dependent fields $\mathbf{u}_a(x)$, defined as wavelet-coefficients of the velocity field $\mathbf{u}$ taken at point $x$ with the resolution $a$. Applying quantum field theory approach of stochastic hydrodynamics to the generating functional of random fields $\mathbf{u}_a(x)$, we have shown the  velocity field correlators 
$\langle \mathbf{u}_{a_1}(x_1)\ldots \mathbf{u}_{a_n}(x_n)\rangle$ 
to be finite by construction for the random stirring force acting at prescribed large scale $L$. 
The study is performed in $d=3$ dimension. 
Since there are no divergences, regularization is not required, and the renormalization group invariance becomes merely a symmetry that relates velocity fluctuations of different scales in terms of the Kolmogorov-Richardson 
picture of turbulence development. 
The integration over the scale arguments is performed from the external scale 
$L$ down to the observation scale $A$, which lies in Kolmogorov range 
$l \ll A \ll L$. Our oversimplified model is full dissipative: interaction between 
scales is  provided 
only locally by the gradient vertex $(\vu\nabla) \vu$, neglecting any  
effects or parity violation that might be responsible for energy backscatter.
The  corrections to viscosity and the pair velocity correlator are calculated in one-loop approximation. This gives the dependence of 
turbulent viscosity on observation scale and describes the scale dependence of the velocity field correlations.
\end{abstract}

\maketitle
\section{Introduction}
Statistical theory of hydrodynamic turbulence has a long history, which stems 
from the work of Osborne Reynolds \cite{Re1895}, suggested that turbulence should be described 
statistically, rather than by the equations of laminar hydrodynamics. 
This practically implies statistical averaging over all possible trajectories,
same as the Feynman's functional integral does in quantum field theory. There is
a standard list of topics in turbulence theory: the stability of solutions of
hydrodynamic equations, the instability of laminar flow, the origin of intermittency, and so on. This paper deals with only one of these topics -- the
description of fully developed homogeneous isotropic turbulence in an
incompressible fluid. The detailed description of the problem can be
found in classical monographs \cite{MonYag,McComb1992,Frisch1995}.

Turbulent flows occurring in various liquids and gasses at high Reynolds numbers 
reveal a number of general aspects: cascade of energy, scaling behavior of correlation functions, statistical correlation laws. However, the prediction of characteristic features of turbulent flow using basic equations of 
fluid dynamics still remains a challenge.

Starting from  phenomenological theory of Kolmogorov-Obukhov  \cite{K41a,K41b,K1962,ob1962}, derived from simple dimensional consideration, it is well known that the probability distribution function of the velocity field fluctuations in fully developed turbulence is 
determined by two basic scales: the microscopic energy dissipation scale $l$ and the macroscopic integral scale $L$, where the energy is injected into the fluid; and also by the energy dissipation rate per unit of mass $\varepsilon$.
The experimental measurements show systematic deviations from Kolmogorov scaling for higher order velocity correlation functions \cite{McComb1992,Frisch1995,Huang2010}. This should be explained from microscopic principles. 

In this paper we consider the Navier-Stokes equation  
\begin{equation}
\frac{\partial \vu}{\partial t}+ (\vu\cdot\nabla) \vu = \nu_0 \Delta \vu
                    - \nabla p  
 + \vf(t,\vx),
\label{nse}
\end{equation}
where $\vu(t,\vx)$ is velocity field, $\nu_0$ is the kinematic viscosity of the fluid, and $p$ is 
the pressure, describing an incompressible fluid stirred by random force $f(t,\vx)$.

The observation scale $a$ is not present in the micromodel \eqref{nse} explicitly: 
the fields $\vu$ and $p$ are square-integrable functions formally defined in 
each point ($t,\vx$). However, we know that the Eulerian velocity 
$\vu(t,\vx)$ is meaningful only in case it is the velocity of fluid averaged over a 
volume $\delta^d$, where $\delta$ is not less than a mean free path, to allow 
for hydrodynamic approximation. So, to introduce the {\em resolution} into 
mathematical consideration of a fluid dynamics problem as an extra coordinate 
$x \to (a,x)$ we extend the space of square-integrable functions $\mathrm{L}^2(\R^{d+1})$ 
by means of continuous wavelet transform performed in a spatial variable $\vx$. 
This trick is quite common in the numerical studies of turbulence \cite{MenSre1987,Farge1992,Yoshimatsu2009}.

The new feature of this paper is that we combine wavelet transform with the 
methods of quantum field theory, including the method of renormalization 
group, to study the statistical momenta of turbulent velocity fields 
at different scales.

The remainder of this paper is organized as follows. In {\sl Section \ref{qft:sec}} we review the quantum field theory approach to fully developed homogeneous isotropic turbulence. In {\sl Section \ref{wt:sec}} we reformulate 
the quantum field theory approach in terms of scale-dependent velocity fields $\mathbf{u}_a(t,\mathbf{x})$. 
The dependence of turbulent viscosity $\nu_a(k)$, which affects the scale components  ${\mathbf{u}}_a(k)$,
calculated in the developed framework, is presented in {\sl Section \ref{rg:sec}}.
Having calculated  the second order statistical momenta $\bra u_a(\omega,\vk) u_{a'}(\omega',\vk') \ket$ in one-loop approximation, in {\sl Section \ref{c2:sec} } we present the energy spectrum of the turbulent velocity field fluctuations as a function of dimensionless observation scale $\xi=A/L$. In {\sl Conclusion} 
we summarize the applicability of our theoretical results to the studies of turbulence. Technical details of 
calculations are presented in {\sl Appendix}.

\section{Quantum field theory approach to statistical hydrodynamics \label{qft:sec}}
Similarity between hydrodynamic turbulence and critical phenomena \cite{WK1974,Vasiliev1998e} made different authors to cast the turbulent velocity field generating functional in a form of quantum field theory, and then 
apply renormalization group  technique, see, e.g.,    \cite{FNS1977,AVP1983,YO1986,YakOrs1986,AAV1996}. 
This approach is valid for the stirred Navier-Stokes description of hydrodynamic turbulence  \eqref{nse}, supplied by the equality condition between the energy injection by random force and the viscous energy 
dissipation.

The generating functional of the velocity field can be written in the form:
\begin{equation}
G[A] = e^{W[A]} = \int \exp\left(S[\Phi] + \int d^d\vx dt A\Phi  \right) \cD \Phi,
\label{gfa}
\end{equation}
where the field $\Phi = (u,u')$ is the doublet of the Eulerian velocity 
field $u(t,\vx)$ and the Martin-Sigia-Rose auxiliary field $u'(t,\vx)$, introduced to exponentiate the delta-function of the equations of motion \cite{MSR1973} (with functional Jacobian of the equations of motion with respect to velocity field being dropped due to appropriate redefinition of the Green functions \cite{AVP1983}, or using ghost fields \cite{AMP1990}). The argument $A(t,\vx)\equiv (A_u,A_{u'})$ is an arbitrary functional source. The ''action'' functional itself takes the form
\begin{equation}
S[\Phi]= \frac{1}{2}\int u' \hat{D}_0 u'  + \int u' [ -\partial_t u + \nu_0 
\Delta u -(u\cdot\nabla)u], \label{act1}
\end{equation}  
where $\hat{D}_0(x-x')=\langle f(x) f(x')\rangle$ is the random force correlator. Integration over the space-time arguments $x\equiv(t,\vx)$ is tacitly assumed. The pressure term is eliminated from the field theory (\ref{gfa},\ref{act1}) by 
the imposed incompressibility conditions $\nabla \cdot \vu = \nabla\cdot \vu'=\nabla\cdot \vf = 0$. 
The incompressibility  is ensured by multiplication of all lines of the 
Feynman graphs of the field theory \eqref{gfa} by the 
orthogonal projector
$$
P_{ij}(\vk) = \delta_{ij} - \frac{k_ik_j}{\vk^2},
$$
where $\vk$ is the momentum of the line.

The perturbation expansion is performed by separating the action \eqref{act1} 
into a free quadratic part  $S_0[\Phi]$ and the cubic interaction term $V[\Phi]$:
\begin{align*}
S_0[\Phi]= \frac{\Phi K \Phi}{2},& & 
K = \begin{pmatrix}
0 & \partial_t + \hat{L} \cr
-\partial_t + \hat{L} & \hat{D}_0 
\end{pmatrix},\\ \hat{L}= \nu_0\Delta,& &  
V[\Phi]= -\frac{1}{2}{u_i}'[\delta_{ik}\nabla_j+\delta_{ij}\nabla_k]u_j u_k .
\end{align*}
The inverse of the matrix $K$ is the matrix of bare propagators. The potential 
$V[\Phi]$ gives the interaction vertex 
$$
v_{ijk} = -\frac{1}{2}[\delta_{ik}\nabla_j+\delta_{ij}\nabla_k].
$$

Within the model \eqref{nse} the pumping power is related 
to the spectral power of the stirring force $d_F(k)$. For the stationary 
isotropic turbulence, stirred by random force $f$, assumed to be delta-correlated 
in time
\begin{equation}
\bra f_i(t,\vx)f_j(t',\vx')\ket = \delta(t-t') \int 
\frac{d\vk}{(2\pi)^d}P_{ij}(\vk) d_F(\vk) e^{\imath\vk(\vx-\vx')},
\label{df}
\end{equation}
the equality of energy injection by random force $f$ to the viscous energy dissipation per unit of mass $\varepsilon$, 
according to Kolmogorov hypotheses \cite{K41a,K41b,K41c}, gives
$$
\varepsilon = \frac{d-1}{2(2\pi)^d} \int d\vk d_F(\vk).
$$
In this paper we are concerned 
with the dimension $d\!=\!3$ for isotropic homogeneous hydrodynamic turbulence.

The field theory of fields $\Phi(x)\in L^2(\R^{d+1})$ determined by 
the action functional \eqref{act1} is UV divergent 
\cite{YO1986,Vasiliev1998e}. To derive quantitative predictions for 
correlation functions the theory should be renormalized using the standard 
method of $\epsilon$-expansion, used in quantum field theory and the 
theory of phase transitions \cite{WK1974,HKN2018}. The difference from 
standard quantum field theory renormalization consists in the role of 
$\epsilon$: in hydrodynamic theory it turns to be the spectral parameter 
of the stirring force rather than the deviation from the dimension of the 
space-time. The choice of the correlator $\hat{D}_0(x-x')$ for the stirring force $f$ in the Navier-Stokes equation \eqref{nse} is a long-standing problem, 
having been discussed at least since \cite{FNS1977}. Most of the papers
exploiting quantum field theory approach to turbulence use stirring force
of IR-type, 
i.e. that concentrated on large scales. This corresponds to shaking the 
''container with turbulence'' as a whole \cite{FNS1977}, although a UV-type 
noise can be also introduced by ''statistical filtering'' procedure of G.Eyink 
\cite{Eyink1996}, which separates fluctuations into large-scale and small-scale 
parts in a way somehow similar to the discrete wavelet transform.  

The main requirements for the stirring force $f$ are the adequate 
description of the large scale behavior of the turbulence and the compatibility 
with the renormalization procedure. A simple power-law choice 
$$
d_F(\vk) = D_0 |\vk|^{4-d-2\epsilon}
$$ 
in \eqref{df} will suffice these requirements at ''realistic'' value of
$\epsilon=2$, which makes the dimension of the constant $D_0$ equal to that 
of the mean energy dissipation per unit of mass $\varepsilon$. The correlator 
$d_F(\vk)$ can be generalized to $|\vk|^{4-d-2\epsilon} h(m/k)$, where $h$ is a 
certain fairly smooth function with $h(0)=1$ \citep{AAV1996}:
\begin{align*}
\bra \tilde{f}_i(t,\vk) \tilde{f}_j(t',\vk')\ket=\delta(t-t')P_{ij}(\vk)
(2\pi)^d \delta^d(\vk+\vk') \\
\times D_0 |\vk|^{4-d-2\epsilon} h(m/k) 
\end{align*}

For convenience of $\epsilon$-expansion the formal expansion parameter 
$g_0=D_0/\nu_0^3$ is introduced. The renormalization procedure looks as 
follows. The original action \eqref{act1}, which depends on two parameters 
$(g_0,\nu_0)$, is declared a ''bare'' action, which yields divergences in the 
perturbation series for the velocity correlators. For realistic value of the 
space dimension $d=3$, the new renormalized action
\begin{equation}
S_R[\Phi]= \frac{1}{2}\int u' \hat{D} u'  + \int u' [ -\partial_t u + Z_\nu \nu 
\Delta u -(u\cdot\nabla)u] \label{actR}
\end{equation}
is derived from the bare action by means of multiplicative renormalization 
$$
\nu_0 = \nu Z_\nu, \quad D = D_0.
$$
The renormalization constant $Z_\nu$, which might be formally infinite, is 
chosen so that it adsorbs the divergences, emerging as poles in $\epsilon$ in 
the perturbation expansion of the velocity field correlator. The renormalized 
parameters $(g,\nu)$ are declared the {\em actual} parameters of the 
perturbation expansion, so that all poles in $\epsilon$ are subtracted from 
the perturbation expansion keeping its finite part intact.

The goal of renormalization procedure is to eliminate the divergences 
appearing in the velocity field correlators. The finite part of the renormalization 
constant $Z_\nu$ is not fixed by this procedure, and therefore may be 
scheme-dependent. The most convenient is the minimal subtraction (MS) scheme 
\cite{Collins1984}. To keep the renormalized coupling constant $g$ 
dimensionless for an arbitrary value of $\epsilon$ an extra parameter $\mu$ of 
the dimension of inverse length is introduced:
$$
g_0 = Z_g g \mu^{2\epsilon},
$$
with the formal coupling constant renormalization defined as $Z_g := Z_\nu^{-3}$ to keep the force correlator invariant under renormalization: 
\begin{equation}
\nu_0^3 g_0 = \nu^3 g \mu^{2\epsilon}. \label{rc0}
\end{equation}

In the space of point-defined functions ($L^2(\R^{d+1})$) the only way 
to reveal the scale dependence of $\bra u(x) u(x')\ket$ is to study the 
dependence of observed velocity correlators on $|\vx-\vx'|$, or alternatively  
on $|\vk|$ in Fourier space. The dependence on the extra parameter of the dimension of length ($1/\mu$) is an artifact of quantum field theoretic 
averaging procedure supplied with subtraction of divergences. The parameter 
$1/\mu$ can be qualitatively understood as the size of the domain over 
which the averaging is performed. But this is a qualitative consideration 
based on similarity of the roles played by the noise dispersion in chaotic systems and the Planck constant in quantum field theory models \cite{Vasiliev1998e}.

To get more physical insight into the problem, we need to use the Kolmogorov 
self-similarity ideas: the turbulence measured at different scales looks 
more or less similar. The space of square-integrable point-defined functions 
is too weak to encompass enough details required for more rigorous
mathematical consideration of self-similarity properties.

At the assumptions on the stirring force mentioned above, the renormalized 
action \eqref{actR} is constructed using a single counterterm, resulting 
in viscosity renormalization $Z_\nu$.  The hydrodynamic field theory $S_R$  
thus has two ''charges'', $g$ and $\nu$, the evolution of which with the 
normalization scale $\mu$ is determined by a single renormalization 
constant $Z_\nu$. In one loop approximation its value is \cite{AVP1983}:
\begin{equation}
Z_\nu = 1 - \frac{ag}{2\epsilon} + O(g^2), \label{zn1}
\end{equation}
with $a=\frac{1}{20\pi^2}$ in $d=3$ dimension.
The $\beta$-function, that determines the evolution of the coupling 
$g$ with the change of scale $\mu$ is derived from the equality \eqref{rc0}:
\begin{equation}
\beta(g) = \left. \mu \frac{\partial}{\partial\mu}\right|_{\nu_0,g_0} g 
= g (-2\epsilon+3\gamma_\nu),
\end{equation}
where $\gamma_\nu := \mu \frac{\partial}{\partial\mu}\ln Z_\nu$. In one-loop 
approximation \eqref{zn1} the $\beta$-function  
\begin{equation}
\beta(g)=-2\epsilon g + 3ag^2 \label{bg1}
\end{equation}
has a IR-stable fixed point $g_* = \frac{2\epsilon}{3a}$, which determines 
the properties of turbulence in large scale asymptotics.

Since the fields ($u,u'$) are not renormalized, any renormalized $n$-point
correlator of velocity field $W^n_R$ is invariant under RG transform: 
\begin{equation}
[\mu \frac{\partial}{\partial\mu} + \beta(g) \frac{\partial}{\partial g} 
- \gamma_\nu(g) \nu \frac{\partial}{\partial\nu}]W^n_R=0. \label{eq43}
\end{equation} 
This means, the statistical momenta can depend only on invariant 
charges $\bar{g},\bar{\nu}$ -- the first integrals of the RG equation 
\eqref{eq43} normalized so that 
$
\bar{g}(s=1,g)=g
$
at the normalization scale, $s=k/\mu$.

The dependence of the invariant 
charge $\bar{g}$ on the invariant scale $s=k/\mu$ is implicitly 
given by the integral of the inverse $\beta$-function:
\begin{equation}
\ln s = \int_g^{\bar{g}} \frac{dx}{\beta(x)}.
\end{equation}
The invariant viscosity $\bar{\nu} = \bar{\nu}(s,g)$ is the second invariant 
of RG equation \eqref{eq43}:
\begin{align}\nonumber 
\bar{\nu} &=& \nu \exp\left[
\int_{\bar{g}}^g \gamma_\nu(x) \frac{dx}{\beta(x)} 
\right] 
= \left( \frac{g\nu^3}{\bar{g}s^{2\epsilon}}
\right)^{1/3} =
 \left(\frac{g_0\nu^3_0}{\bar{g}k^{2\epsilon}}\right)^{1/3}.
\end{align}
At the presence of the IR-stable fixed point $\beta(g_*)=0$ in \eqref{bg1}, the turbulence 
behavior at large scales is determined by the value of invariant viscosity 
at fixed point $g_*$:
\begin{equation}
\bar{\nu}_*(k) = \nu_0 \left(\frac{g_0}{g_*}\right)^{1/3} k^{-\frac{2\epsilon}{3}}. \label{nu-fp}
\end{equation}
The pair correlator of velocity field $C=\bra uu \ket$ that satisfies 
RG equation \eqref{eq43} has the form 
$$
C = \bar{\nu}^2 k^{-d} R(\bar{g},\bar{z}),
$$
where $R(\cdot)$ is some function of the invariant coupling constant $\bar{g}$ 
and the invariant frequency $\bar{z}=\frac{\omega}{\bar{\nu}k^2}$. (More details 
and the incorporation of IR scale parameter $m$ into consideration can be found 
in \cite{AAV1996}.)
The equal-time pair correlator is obtained by integrating $C$ over the
frequency argument:
\begin{equation}
C_{st} = \int C \frac{d\omega}{2\pi} = \bar{\nu}^2 k^{2-d} R(\bar{g}).
\label{cst}
\end{equation} 
The substitution of the viscosity \eqref{nu-fp} into \eqref{cst} at 
$\epsilon=2,d=3$ yields the IR asymptotics of the Kolmogorov type: 
$
C_{st} \propto k^{-\frac{11}{3}}.
$
Further discussion on anomalous scaling, different from the Kolmogorov regime, the 
effects of anisotropy, compressibility, and finite correlation time effects can be 
found, e.g., in \cite{AHHJ2003}.

\section{Multiscale theory of turbulence in wavelet basis \label{wt:sec}} 
Kolmogorov (K41) hypotheses \cite{K41a} assume statistically homogeneous and isotropic turbulence. This justifies the evaluation of velocity field correlations in wavenumber space, but does not provide any rigorous mathematical definition of the 
''fluctuation of scale $a$''. They are tacitly assumed in the literature as 
Fourier components of velocity field with wave numbers equal to the inverse 
scale: $|\vk|\!\approx\!\frac{2\pi}{a}$. Such nonlocal definition meets global characteristics of the homogeneous isotropic turbulence, but is hardly
applicable to nonlinear phenomena such as coherent structure formation.

To analyze the {\em local} properties of turbulent velocity field at a 
given scale $a$, same as in quantum field theory \cite{Altaisky2010PRD,AK2013}, the wavelet decomposition 
$\vu(t,\vx,\cdot)\to \vu_a(t,\vx,\cdot)$ was performed by many authors, see e.g., \cite{Zimin1981,Meneveau1991,YO1991,VerFri1991,Farge1992,Ast96e}. Among those, the 
wavelet transform was applied to the iterative solution of the stochastic Navier-Stokes equation \cite{AltDAN2006}.     

To perform wavelet decomposition of the velocity field  we 
need some aperture function $g(\vx)\in {\rm L}^2(\R^d)$, called a {\em basic wavelet}, which satisfies an admissibility condition 
\begin{equation}
C_g = \int_0^\infty |\tilde{g}(a)|^2 \frac{da}{a}<\infty,
\label{adc}
\end{equation}
 so that the 
 original (''no-scale'') field $u(t,\vx,\cdot)$ can be reconstructed from the set of its wavelet coefficients $u_a(\vb,\cdot)$:
\begin{align}\nonumber 
u(\vx,\cdot) &=& \frac{1}{C_g}\int_0^\infty \frac{da}{a}\int_{\R^d} \frac{1}{a^d}g
\left(\frac{\vx-\vb}{a} \right)u_a(\vb,\cdot) d^d\vb, \\
u_a(\vb,\cdot) &=& \int_{\R^d} \frac{1}{a^d}\bar{g}
\left(\frac{\vx-\vb}{a} \right) u(\vx,\cdot) d^d\vx. \label{cwt} 
\end{align}
The wavelet coefficients 
$u_a(\vb,\cdot)$ can be considered as the scale components of the velocity
field $u$ measured with the aperture function $g$.
To keep the fields $u$ and $u_a$ the same physical dimension the ${\rm L}^1$ norm is used in wavelet transform \eqref{cwt}  instead of the 
traditional ${\rm L}^2$ norm \cite{HM1998,Altaisky2010PRD}.

In contrast to ''statistical filtering'' procedure of \cite{Eyink1996}, which is also given by convolution with a filtering function $G_l(x):=l^{-d}G(x/l)$, the continuous wavelet transform \eqref{cwt} is {\em invertible}.  
Usage of basic wavelet that satisfies the admissibility condition \eqref{adc} 
makes our theory significantly different from ''statistical filtering''. The 
difference is briefly as follows. The statistical filtering operator $G_l$ 
projects the velocity field onto the space of functions $\cF_k$ with wave vectors 
less or equal to a given value $k \sim 1/l$. Thus $G_l$ is a low-pass filter:
$$
v_l(x):= G_l * v(x).
$$
The projection $\mathrm{L}^2(\R^d) \stackrel{G_l}{\to} \cF_k$ is a homomorphism.  The details lost by this projection are given 
by the high-pass filter $H_l$, so that 
$$
\hat{G}_l(k) + \hat{H}_l(k)=1.$$
Statistical filtering applies the low-pass and high-pass filters only once, 
and then treat large-scale and small-scale modes \cite{Eyink1996}. 
The Kadanoff blocking procedure \cite{Kadanoff1966} applies it sequentially to coarser and coarser 
scale, each time increasing the size of the block by an integer factor and loosing 
some details on each step. The renormalization group can do it gradually, integrating over the difference space
\begin{equation}
D_{k,\Delta k}:= \cF_k \setminus \cF_{k-\Delta k} \label{Dk:def}
\end{equation} 
on each step.
Since $\ldots \cF_{k-2\Delta k} \subset \cF_{k-\Delta k} \subset \cF_{k}$, 
the spaces \eqref{Dk:def} allow for an evident decomposition 
\begin{equation}
\cF_k = D_{k,\Delta k} \oplus D_{k-\Delta k,\Delta k} \oplus D_{k-2\Delta k,\Delta k} \oplus \ldots. \label{fkd}
\end{equation}
To study the behavior of a function on a ladder of scales $k,k-\Delta k,k-2\Delta k,\ldots$ it is sufficient to project it onto a set of 
the difference subspaces \eqref{fkd}, with no need to keep the whole 
set $\{ \cF_k \}_k$. 
The decomposition \eqref{fkd} is exactly what wavelet transform does, if  
discretized in an orthogonal basis, see, e.g, \cite{Daub10}. 
So, in our approach we separately treat the fluctuations concentrated {\em near} each given scale $\{ u_l(x)\}_l$, rather than all fluctuations concentrated above the given scale $l$, as G.Eyink does. Thus integrating from external size 
of the system $L$ down to the observation scale $A$ we can reconstruct velocity field 
in the sense of \eqref{fkd}. No need to say that the Gaussian filtering 
cannot be used for wavelet decomposition for it does not satisfy the 
admissibility condition \eqref{adc}, and therefore the original 
function $v(x)$ cannot be uniquely reconstructed from the set of its coefficients $v_l(x)$.

Referring the reader to general textbooks in wavelet analysis \cite{Daub10,Chui1992} for more details 
on the continuous wavelet transform \eqref{cwt}, we assume for simplicity 
the basic wavelet $g(\vx)$ to be isotropic function of $\vx$, having fairly good localization properties; it may be a derivative of Gaussian, for instance, \cite{Farge1992,Frisch1995}. In Fourier space the convolution becomes a product: $\tilde{u}_a(\vk) = \bar{\tilde{g}}(a\vk) \tilde{u}(\vk)$.  

The stirring force can be represented by its scale components $f_a$, Gaussian random functions with zero mean, concentrated at a fixed large 
scale $L$. The correlator of the stirring force scale components can be taken 
in the form 
\begin{align} \nonumber 
\langle \tilde{f}_{ai}(t,\vk) \tilde{f}_{a'j}(t',\vk') \rangle &=&
\delta(t-t') P_{ij} (\vk) g_0 \nu_0^3 C_g (2\pi)^d  \times \\
&\times& \delta^d(\vk+\vk') a \delta (a-a') \delta (a-L), \label{fcor} 
\end{align}
where $g_0$ is formal dimensionless strength of forcing.
The random force \eqref{fcor} in our approach simulates random 
initial/boundary conditions, and also the effect of external forces 
injecting energy from large scales comparable to the size of the turbulent 
domain. The delta-function-type correlator is of course an approximation
that enables 
analytical calculations of the diagram expansion. A random force concentrated 
on a narrow range of gross scales would be more realistic, but is hard for 
analytical calculations. We follow the Kolmogorov-Richardson scenario of 
turbulence development: the kinetic energy injected at large scale by 
external forcing is transferred to smaller and smaller scales until it reaches 
the scale $l$, where it is dissipated by viscosity. 
We do not consider the effect of small 
scales close to Kolmogorov dissipative length $l$ on the dynamics 
of larger eddies. In this respect there is a difference from the 
 approach \cite{Eyink1996}, where the random force is split into a large 
scale part $\bar{f}$ and the high-frequency noise acting on molecular 
scales.

In our model the molecular noise and the subgrid effects below 
the observation scale $A$ are neglected for they do not seriously 
affect the large scale motion.
The reason is that our consideration is concerned with a homogeneous isotropic 
stationary turbulence with no parity violation, i.e, with the mean helicity assumed to be zero $\bra \vu \cdot 
\mathrm{curl}\ \vu \ket=0$.  We have only two inviscid invariants: the kinetic 
energy and the helicity. The inverse 
energy cascade can be induced by the presence of extra topological 
invariant -- the conservation of enstrophy the  $Z=\frac{1}{2}\int\mathrm{curl}^2\vu$
\cite{MonYag,Frisch1995}. In a fully three-dimensional turbulence the enstrophy is not conserved and 
the inverse energy cascade is not significant in the inertial range 
of scales. In our simple model we consider isotropic basic wavelets.  
The  quasi-two-dimensional turbulence, with one dimension being much less than the others, causing inverse energy cascade \cite{CMV2010}, can be hardly handled analytically in our framework. Our consideration may be not true for helical turbulence, but this is not the subject of this paper, keeping it for future research.

The basic objects of our model are the correlation functions of the  of the velocity field scale 
components $\langle \Phi_{a_1}(x_1)\ldots \Phi_{a_n}(x_n)\rangle$. They can be evaluated from the generating functional 
\begin{equation}
G[A] = e^{W[A]} = \int \cD \Phi_a(x) e^{S[\Phi_a] + \int  \frac{dx da}{a} A_a(x)\Phi_a(x)}, 
\label{gfaa}
\end{equation} 
which is different from its classical counterpart \eqref{gfa} only by making the integration measure $dx\equiv dtd^d\vx$ into $dt\frac{d^d\vx da}{a}$ for each 
space-time argument, and substituting the interaction vertex $v_{ijk}$ by its 
wavelet transform. The substitution of wavelet transform \eqref{cwt} into the action functional 
\eqref{act1} yields the action functional of the scale-dependent fields
\begin{align} \nonumber 
S[\Phi_a] &=& \frac{1}{2}\int \frac{dx da}{a}\frac{dx' da'}{a'}u_{a}'(x)
D_{aa'}(x-x') u_{a'}'(x')  + \\ &+& \int \frac{dx da}{a}
u_{a}'(x)[-\partial_t u_a(x) + \nu_0 \Delta u_a(x) 
+ V_a[u]], \label{act2}
\end{align}
where $V_a[u]$ is an integral nonlinear operator, obtained by wavelet transform of the cubic interaction term $u' u \nabla u$.

We use the first derivative of the Gaussian as a basic wavelet 
$g$. The equality between the energy injection and the energy dissipation 
then defines the bare coupling constant $g_0$:
\begin{equation}
\varepsilon = \frac{g_0\nu_0^3}{L^4} \frac{3}{8\pi^{3/2}}. \label{wg1}
\end{equation}
See {\sl Appendix \ref{asf:sec}}  for details. 

The functional derivatives are taken with respect to the formal source $A_a(t,\vx)$:
$$
\langle \Phi_{a_1}(x_1)\ldots \Phi_{a_n}(x_n)\rangle_c = \left.
\frac{\delta^n W[A]}{\delta A_{a_1}(x_1)\ldots \delta A_{a_n}(x_n)}
\right|_{A=0}.
$$ 
Integration over all scale arguments $\int_0^\infty \frac{da_i}{a_i}$ would certainly drive us back to the known divergent description of fully developed turbulence, in case we substitute the force correlator \eqref{fcor} by wavelet image of a wide-band correlator  power-law correlator \cite{AAV1999}.

Taking into account that statistical properties of  fully developed turbulence are determined by the energy flux from large scales to small 
scales, we  apply the following rule for the calculation of any 
Feynman graph for the correlation functions   $\langle \Phi_{a_1}(x_1)\ldots \Phi_{a_n}(x_n)\rangle$. Let $A=\min(a_1,\ldots,a_n)$, then the integration 
in all internal lines is to be performed within the range $\int_A^\infty \frac{da_i}{a_i}$.
The theory defined in this way is finite by construction \cite{Alt2002G24J,Altaisky2010PRD,AK2013}.  In contrast to standard means of regularization, 
such as introduction of cutoff momenta, our method 
provides an exact conservation of momentum in each vertex.

\section{One-loop corrections to viscosity \label{rg:sec}}
\begin{figure}[t]
\centering \includegraphics[width=4cm]{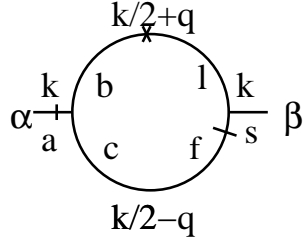}
\caption{One-loop contribution to the Green function. Primed lines denote 
the $u'$ fields incident to the vertices. Crossed line denotes the stirring 
force correlators. Latin letters indicate vector indices, Greek letters stand for the scale arguments of external lines}
\label{gf1l:pic}
\end{figure}

Bare Green function of the  field theory \eqref{gfa} are determined by the linear part of the Navier-Stokes operator. 
In multiscale theory the bare response function between the scales $\alpha$ and $\beta$ is obtained by multiplication of ordinary one by the wavelet factors: 
$$
G^{(0)}_{\alpha,i,\beta,j}(k) = \frac{\tilde{g}(\alpha\vk) \overline{\tilde{g}}(\beta\vk)}{-\imath k_0 + \nu_0 \vk^2}P_{ij}(\vk).
$$
One loop contribution to this Green function is graphically shown in Fig.~\ref{gf1l:pic}.
For the simplicity of calculations, same as in \cite{Altaisky2010PRD}, we have chosen the first derivative of the Gaussian as the basic wavelet. Its shape naturally resembles a localized wave. Its Fourier transform is $\tilde{g}_1(k)=-\imath k e^{-\frac{k^2}{2}}$, 
with $C_{g_1}\!=\!\frac{1}{2}$. Due to the limited range of integration over the scale 
variables $\int_A^\infty \frac{da_i}{a_i}$, bounded from below, all  
internal lines will contribute a cutoff factor $f_g^2(Ak)$, where $k$ is 
the momentum of the line, and $f_g(x) = \frac{1}{C_g} \int_x^\infty \frac{|\tilde{g}(a)|^2}{a}da$; for $g_1$ wavelet cutoff factor is $f_{g_1}(x)=e^{-x^2}$, see 
\cite{Altaisky2010PRD,AK2013} for details. 

To calculate one-loop contribution to viscosity coming from large scale turbulent 
pulsations we consider 1PI diagrams for the two-point vertex function $\Gamma^{(2)}$:
\begin{equation}
\Gamma^{(2)} = \Gamma^{(2)}_{bare} + \Sigma_{\alpha\beta},
\end{equation}
where $-\Sigma_{\alpha\beta}$ is the value of the diagram shown in Fig.~\ref{gf1l:pic}. The explicit equation for the ''self-energy'' $\Sigma_{\alpha\beta}(k)$ is (the details are presented in the {\em Appendix \ref{agf2:sec}}):
\begin{align}\nonumber 
\Sigma_{as} &=& -\nu_0  g_0  kL\int_0^\infty \frac{y^2 dy}{(2\pi)^2} 
e^{-(kL)^2 (1+4\xi^2)\left(\frac{1}{4} + y^2 \right)} \times \\
&\times& \int_0^\pi d\theta\sin\theta  
L_{as}(k,p^+,p^-) 
\frac{e^{-(kL)^2  y \cos \theta} }{\frac{1}{4}+y^2 -\imath
\frac{\omega}{2\nu_0 \vk^2}
}
, \label{se}
\end{align}
where we have introduced a dimensionless scale $\xi=\frac{A}{L}$ and 
the dimensionless momentum $\vy = \frac{\vq}{|\vk|}$ for integration in  
$\R^3$. 
The one-loop tensor structure of the diagram shown in Fig.~\ref{gf1l:pic} is 
\begin{align} \nonumber 
L_{as}(k,p^+,p^-) &=& \frac{\delta_{as}}{4}
\left[
\frac{(p^+  k) (p^+  p^-)}{{p^+}^2} - (k p^-)
\right]   \\ 
\nonumber &+& \frac{p_a^+ p_s^-}{2} 
\left[
\frac{(p^+ k)}{{p^+}^2} - \frac{(p^- k)(p^+ p^-)}{{p^-}^2 {p^+}^2}
\right] \\ 
\nonumber &+&p^-_a p^-_s 
\left[
\frac{(kp^-)}{{p^-}^2}-\frac{(p^+k)(p^+p^-)}{2{p^-}^2{p^+}^2}
\right] \\ 
&-&\frac{k_a p^-_s}{2}
+p^+_ak_s \frac{p^+p^-}{4{p^+}^2} - \frac{p^-_a k_s}{4}, \label{ts1}
\end{align}
with $p^\pm = \frac{k}{2} \pm q$, and all scalar products taken in $\R^3$. For the isotropic turbulence, the tensor structure of the ''self-energy'' diagram \eqref{se} may depend 
only on the direction of vector $\vk$. It can be written in the form 
\begin{equation}
\Sigma_{as} = \nu_0 g_0 \Sigma^\delta k^2  \left(\delta_{as} - \frac{k_a k_s}{k^2}\right) + \nu_0 g_0\Sigma^\lambda k_a k_s .
\end{equation}
After standard algebraic manipulations this gives 
\begin{widetext}
\begin{align*}
\Sigma^\delta  = \frac{kL}{128 C_g} \int_0^\infty \frac{y^2 dy}{(2\pi)^2}
\frac{e^{-(kL)^2(1+4\xi^2) \left(\frac{1}{4}+y^2 \right)}}{\frac{1}{4}+y^2-\imath
\frac{\omega}{2\nu_0 \vk^2}
} \int_{-1}^{1}d\mu
\frac{(1-\mu^2)(8\mu^2y^2+\mu(8y^3-10y)+4y^2+1)}{\left(\frac{1}{4y}+y-\mu \right)\left(\frac{1}{4y}+y+\mu \right) } e^{-(kL)^2 y\mu},
\end{align*} 
\end{widetext}
where $\mu=\cos\theta$, with $\theta$ being the polar angle between $\vk$ and $\vq$.

In our model the ''self-energy'' contribution to viscosity 
is finite by construction. It determines the relation between the viscosity  measured at certain reference scale, and the viscosity  at the observation scale $A$. 
Following the Kolmogorov-Richardson scenario of turbulence development \cite{Richardson1922}, we sum up all 
fluctuations from the large stirring scale $L$ up to the measuring scale $A$, 
where $\xi = A/L \ll 1$, but still much above the Kolmogorov scale $A \gg l = (\nu^3/\varepsilon)^{1/4}$. Similar settings were used for renormalization group studies of the general 
case of helical turbulence, where the dependence of energy spectra on observation scale 
was shown to be significant only on the edges of of the energy-containing range 
by means of affecting the stability of RG fixed points \cite{Ch2006jpa,Ch2006pre}.

Measuring all lengths in units of $L$  we rewrite renormalization of viscosity in the form  
\begin{equation}
\nu(\xi) = \nu_L Z_\nu,\quad Z_\nu= 1-g_L\Sigma^\delta(\xi).
\label{nua}
\end{equation}
Equation \eqref{nua} works fairly well if the difference between 
the observation scale $A$ and the stirring scale $L$ is not too big,
otherwise we need to solve RG equations to 
determine $g_L$ and $\nu_L$ as functions of the microscopic parameters 
$g_0$ and $\nu_0$.

The RG equations may be obtained by iterating the equation \eqref{nua} 
over the set of scales $$1=\xi_L>\xi_{L-1}>\xi_{L-2}>\ldots>\xi_0=\frac{l}{L}, \xi_k = \xi_0 \delta^k, \delta>1.$$
In continuous limit this leads to the RG equation 
\begin{equation}
\frac{d\ln\nu}{d\ln\xi}=g(\xi) \frac{\Sigma(\xi)}{\ln\delta},
\label{rge-nu}
\end{equation}
see {\sl Appendix \ref{anu:sec}} for the derivation.

Evolution of the formal coupling constant $g(\xi)$ is determined 
by the scale corrections to the stirring force correlator 
$$D(\xi)=\frac{g(\xi)\nu^3(\xi)}{L}.$$ In one-loop approximation, 
 the renormalization of stirring force correlator is given by 
\begin{equation}
\frac{d\ln D}{d\ln\xi} = - \frac{K(\xi)}{\ln\delta},
\label{rge-D}
\end{equation}
where $K(\xi)$ is one-loop contribution to the stirring 
force correlator, see the {\em Appendix } \ref{adf:sec} for details.
Making use of RG equations (\ref{rge-nu},\ref{rge-D}), and since 
$\ln g(\xi) = \ln D(\xi) - \ln L - 3\ln \nu(\xi)$,
we get the RG equation for 
the formal coupling constant $g(\xi)$:
\begin{equation}
\frac{d\ln g}{d\ln\xi} = - \frac{K(\xi)}{\ln\delta} -3 g(\xi) \frac{\Sigma(\xi)}{\ln \delta},
\label{rge-g}
\end{equation}
which has the solution 
\begin{equation}
g(\xi) = \frac{g_L e^{\int_\xi^1 \frac{d\eta}{\eta}K(\eta) }}{1-3g_L\int_\xi^1 \frac{d\xi'}{\xi'}\Sigma(\xi') 
e^{\int_{\xi'}^1 \frac{d\eta}{\eta}K(\eta)}
}, \label{gxi}
\end{equation}
where we have set $\ln\delta=1$.

In view of $K(\xi) \ll 1$, the value of $D(\xi)$ is utmost scale-invariant
and we can use the equality $$ g(\xi) \nu^3(\xi) = g_0 \nu_0^3 = g_L \nu_L^3 $$ to 
evaluate $\nu(\xi)$ for the known values of $g(\xi)$.

\begin{figure}[t]
\centering \includegraphics[width=2.8in]{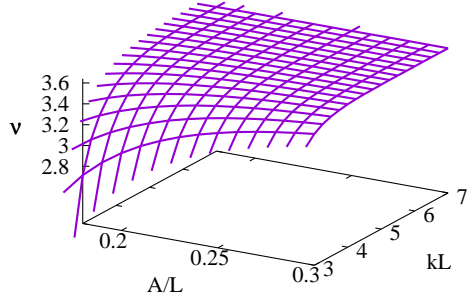} 
\caption{The dependence of turbulent viscosity $\nu_A(k)$ on the observation scale $\xi=A/L$ and the dimensionless momentum $x=kL$. Calculations were performed in the IR region using the equation \eqref{gxi} at the assumption of invariant $g_A \nu_A^3 = g_L \nu_L^3$, i.e., small $K(\xi)$, and normalization momentum $x_*=4\pi$ used to evaluate 
$g_L$, in accordance to $k=2$ energy injection limit in simulated turbulence from John Hopkins Turbulence Database}
\label{nu2ir:pic}
\end{figure}

The solution of the RG equation \eqref{rge-nu} in the IR region, calculated for a fixed normalization  momentum $x_*=4\pi$, is presented in Fig.~\ref{nu2ir:pic}.
The renormalized viscosity $\nu_\xi(k)$ is a counterpart of renormalized 
viscosity in the action of ordinary theory \eqref{actR}. The difference is 
that $\nu_\xi(k)$ is taken not in IR-stable fixed point, and therefore 
describes the asymptotic behavior of large-scale eddies, but is taken at 
a fixed observation scale $A=\xi L$. In terms of the scale-dependent 
action $S[u_a,u_a']$ $\nu_\xi(k)$ can be understood as a viscosity acting 
on the wavelet-type pulsations of the velocity field $u_a(x)$ measured at 
scale $A$. The wave vector $k$ of such perturbations can take arbitrary 
values. 

In this paper we neither aim to construct a turbulent stress tensor, 
as is presented in statistical filtering theory \cite{Eyink1996} for 
the wave vectors less or equal to $1/a$, nor we construct statistical 
closures for scale-dependent fields $\vu_a(x)$ for it would result in integral 
equations. Instead, since what is really measured in turbulence are the 
$n$-point correlation functions, we consider the Fourier transform of such 
functions $\bra \tilde{\vu}_{a_1}(k_1)\ldots \tilde{\vu}_{a_n}(k_n)\ket$,
where the wave vectors $k_i$ are responsible for the {\em separation} between 
the observation points, while the scale arguments $a_i$ are responsible 
for the observation scales.

\section{Energy spectra \label{c2:sec}} 
The full kinetic energy of homogeneous isotropic turbulence can be expressed
in terms of scale components of the velocity field:
$
E = \frac{1}{2} \int \bra |{\tilde{\vu}}_a(\vk)|^2\ket
\frac{d^dk}{(2\pi)^d}\frac{1}{C_g}\frac{da}{a}
$.
\begin{figure}[ht]
\centering \includegraphics[scale=0.5]{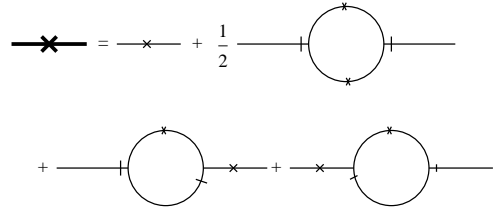}
\caption{Feynman diagrams used for evaluation of the pair correlator of velocity field}
\label{c2:pic}
\end{figure}
We evaluate the the equal-time pair correlator $C(k,\xi)=\bra \tilde u_A(t,\vk) \tilde u_A(t,-\vk) \ket$ of velocity field wavelet 
coefficients according the 
diagram rule shown in Fig.~\ref{c2:pic}.
The first term, the bare correlator integrated over the frequency and two internal scales, is equal to
\begin{equation} 
C^{(0)} 
= \frac{1}{\nu_0k^2}\frac{g_0\nu_0^3}{L} (Lk)^2 e^{-(Lk)^2}|\tilde{g}(Ak)|^2 P_{ik}(\mathbf{k}).  \label{C0}
\end{equation}

The next contribution comes from the symmetric diagram in first line of Fig.~\ref{c2:pic}, integrated over the frequency arguments. This gives
\begin{widetext}
\begin{align}
C^{(2)} = \frac{1}{2} \left( \frac{g_0\nu_0^3}{2C_g} \right)^2
\frac{L^2k^3}{16\nu_0^3}|\tilde{g}(Ak)|^2 f^2(Ak) P_{ik}(\mathbf{k}) 
\int \frac{y^2 dy}{(2\pi)^2}
\frac{ e^{-2(Lk)^2(1+2\xi^2)\left(\frac{1}{4}+y^2\right)}\left(\frac{1}{4}+y^2\right)
}{
4 \left(\frac{1}{4}+y^2 \right) \nu_0 k^2 \left(1+ \frac{1}{2}
\left(\frac{1}{4}+y^2\right)
 \right) 
}
\int d\mu 
\frac{(1-\mu^2)(8\mu^2 y^2+4y^2+1)}{
\left(\frac{1}{4}+y^2-y\mu \right) \left(\frac{1}{4}+y^2+y\mu \right)
},
\end{align}
\end{widetext}
where $f(Ak) = e^{-(Ak)^2}$ are wavelet filters in the legs of diagram. $1/2$ before 
the whole equation is a topological factor.

Two last diagrams in Fig.~\ref{c2:pic} contribute equally to the correlation function. Their joint contribution is
\begin{widetext}
\begin{align}\nonumber 
2C^{(1)} &=& -2 P_{ik}(\mathbf{k})|\tilde{g}(Ak)|^2 f^2(Ak) \frac{g_0\nu_0^3}{C_g L} |\tilde{g}(Lk)|^2 
\frac{g_0 kL}{128C_g} 
\int \frac{y^4 dy}{(2\pi)^2}
\frac{ e^{-(Lk)^2(1+4\xi^2)\left(\frac{1}{4}+y^2\right)}
}{
2(\nu_0 k^2)^2 \left(1+ 2
\left(\frac{1}{4}+y^2\right)
 \right) 
} \times \\ 
&\times& \int d\mu  
\frac{(1-\mu^2)(8\mu^2y^2 +2 \mu y(4y^2-5)+ 4y^2+1)}
{ 
\left(\frac{1}{4}+y^2-y\mu \right) \left(\frac{1}{4}+y^2+y\mu \right)
} 
e^{-(kL)^2 y\mu}.
\end{align}
\end{widetext}
The common sign minus stands for the fact $\Sigma^\delta$ is equal to minus diagram.

The final equation, without common wavelet factor $\tilde{g}(kA) \bar{\tilde{g}}(kA)$ on the legs of each diagram, 
calculated with $g_1$ wavelet is given by:
\begin{widetext}
\begin{align} \nonumber 
& & C(k,\xi) = \frac{g_0\nu_0^3}{\nu_A(k)}L e^{-(Lk)^2} \\
\nonumber &+& \frac{(g_0\nu_0^3)^2}{128} \frac{(Lk)L}{\nu_A^4(k)}e^{-2\xi^2 (Lk)^2} \int_0^\infty \frac{y^2 dy}{(2\pi)^2} \frac{e^{-2(kL)^2(1+2\xi^2)
     \left(\frac{1}{4}+y^2 \right)}}{1 + \frac{1}{2} \left(\frac{1}{4}+y^2 \right)}
     \int_{-1}^1 d\mu \frac{
(1-\mu^2)(8\mu^2 y^2+4y^2+1)     
     }{\left(\frac{1}{4}+y^2-y\mu \right)\left(\frac{1}{4}+y^2+y\mu \right)
     } 
+ \frac{(g_0\nu_0^3)^2}{32} \frac{(Lk)L}{\nu_A^4(k)}\\
&\times& e^{-(Lk)^2(1+2\xi^2)} 
     \int_0^\infty \frac{y^4 dy}{(2\pi)^2} \frac{e^{-(kL)^2(1+4\xi^2)
     \left(\frac{1}{4}+y^2 \right)}}{1 + 2 \left(\frac{1}{4}+y^2 \right)} 
  \int_{-1}^1 d\mu \frac{
(\mu^2-1)(8\mu^2 y^2 +2\mu y (4y^2-5)  +4y^2+1)     
     }{\left(\frac{1}{4}+y^2-y\mu \right)\left(\frac{1}{4}+y^2+y\mu \right)
     }     e^{-(kL)^2 y \mu}. \label{Ck}
\end{align}
\end{widetext}

The coupling $g_0$ is related to the mean energy dissipation rate per unit of mass $\varepsilon$ by the energy balance equation 
$\varepsilon L^4 = g_0 \nu_0^3 \chi[g],$
where $\chi[g]$ is numeric factor, which depends on the shape 
of basic wavelet \eqref{wg1}. For the $g_1$ wavelet used in this paper 
$\chi[g_1]=\frac{3}{8\pi^{3/2}}$.

The obtained function \eqref{Ck} can be used to study the
dependence of the turbulent pulsations energy spectrum 
$E(x,\xi)=4\pi k^2 C(k,\xi)$, which is assumed to be  
Kolmogorov spectrum, if there is no dependence  on the observation scale $\xi=A/L$.

We have compared our results with the $1024^3$ grid simulations of homogeneous isotropic turbulence 
presented in John Hopkins Turbulence Database (JHTDB) and described, e.g., in \cite{Eyink2008}, 
with the following parameters of simulation: cubic domain of size $L=2\pi$, Kolmogorov length $l=0.0028$, dissipation rate $\varepsilon= 0.103$, viscosity $\nu_0=0.000185$. 

In traditional theory of turbulence the dependence of results of measurements, {\sl viz.} 
the fields $v_\delta:= \frac{1}{\delta^d}\int v(x)d^dx$ and their statistical momenta, on the averaging scale $\delta$ is usually considered as an artifact of inappropriate choice: 
if $\delta$ is too small, say is close to the mean free path, we may get out of applicability 
of hydrodynamic approximation; alternatively, if it is too large, that is of the same order as the system size $L$, we get out of the limits of the Kolmogorov theory. Thus the ''legitimate'' choice of observation scale lies deeply inside the Kolmogorov range of scales: $l \ll \delta \ll L$. Experimental processing of turbulence data stepped a little further when 
the wavelet transform was used to study turbulence behavior in ($k,a$) plane \cite{VerFri1991}. The study of distributions in ($k,a$) plane gives more information 
than that in $k$ only: the window width ($a$) for each mode ($k$) may tell whether this 
mode originates from the small or from the large scale dynamics. 

In the limit of large observation scales $a \lesssim L$ there is no need for pulsations 
$u_a(k)$ to obey Kolmogorov's laws. The energy of such pulsations decreases with 
the increase of resolution $a\to L$. The analytic tools, based on continuous wavelet transform, we propose in this paper 
may be useful in analytical computations of correlations of velocity pulsations 
measured at different spatial resolution.   

\begin{figure}[t]
\centering \includegraphics[width=8cm]{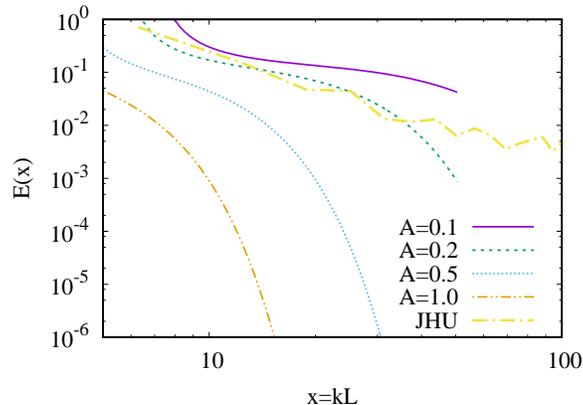}
\caption{One dimensional energy spectra. Calculated for the set of parameters 
of the John Hopkins Turbulence Database  for the isotropic turbulence  grid simulations. The  curves show the spectra for observation scales $A=0.1,0.2,0.5,1.0$. 
The $E(k)$ numerical energy spectrum, obtained for the $128^3$ decimation of the original JHTDB $1024^3$ data, is shown by dashed line. The normalization scale $x_*=4\pi$ corresponds to the $k=2$ energy injection limit. All quantities are normalized to $(2\pi)^3$ volume}
\label{ek:pic}
\end{figure}

As we can see from  
Fig.~\ref{ek:pic}, the slope of the curves $E(k,\xi)$ depends on the observation scale $\xi$.
The curves $A=0.1,0.2$, corresponding to small observation scales, i.e. those more than an order of magnitude less than external scale $L$, have the slopes close to the Kolmogorov 
$k^{-\frac{5}{3}}$ regime. In contrast, the larger observation window $A=0.5,1.0$, i.e. 
only one order less than $L$, results in a steeper falloff of the energy curves.
Same thing happens with the dependence of energy on the dimensionless wave vector 
$x=kL$. Our analyzing wavelet $\tilde{g}(ak) \sim (ak) e^{-\frac{(ak)^2}{2}}$ is mostly sensitive to the wave numbers $k\sim\frac{1}{a}$, hence the observation scale $A\sim 0.1$ 
results in dimensionless wavevectors of the order $x \sim \frac{2\pi}{0.1} \approx 63$, 
and similar for other curves, which qualitatively agrees with that observed in
Fig.~\ref{ek:pic}.

The spectral index is close to the Kolmogorov 
value $-\frac{5}{3}$ for the values of observation scale in the 
middle of the inertial range $\xi_0 \ll  \xi \ll 1,$ but becomes 
steeper when approaching the dissipation scale $\xi_0$. Since our correlation functions 
\eqref{Ck} represent only partial energies of the given scale $\xi=A/L$ fluctuations, only 
the slopes can be compared. The integral over all scales should give a "no-scale" energy spectrum.  
Since we are interested in the dependence of the energy spectra on the observation scale 
$\xi$ in Fig.~\ref{ek:pic} we present the graphs of such spectra and the standard "no-scale" 
spectrum (shown in dashed line) obtained from numerical simulations http://turbulence.pha.jhu.edu

\section{Conclusion}
We conclude, that using the multiscale representation of  fields 
 in field-theoretic  calculations of turbulent velocity correlations we can obtain more  
information on the statistics of turbulent pulsations, than by  
standard spectral methods. Presented formalism is more relevant to experimental 
studies: any measured statistics of turbulent pulsations is 
always obtained at certain observation scale  by averaging velocity 
fluctuations over the measuring volume. This volume should be somehow 
taken into account by theoretical description of turbulence.

It should be noted that our method of renormalization, acting 
from larger scales to smaller scales, ignores any effects of 
inverse energy cascade. This is significant simplification. The authors 
think, however, that a rigorous approach to the effect of small scale 
fluctuations on large scale fluctuations, which remains a chalenging problem, 
would involve more complicated 
description than just the forced Navier-Stokes equation itself. First of 
all the parity breaking effects and the inviscid topological invariants  
should be taken into account \cite{Ch2006pre}. 
The complexity of the problem can be seen for instance from the recent paper \cite{mcy2017}. 
We can hardly foresee 
such theory of an incompressible fluid flow capable of analytic calculations in the nearest 
future. The incompressibility itself is also a simplification. Wavelets might be of some use here as well, but up to the best authors knowledge 
their use in studying inverse energy cascade is limited to numerical simulation 
\cite{Meneveau1991prl}.

From experimental point of view the dependence of the 
energy spectra on the observation scale $A$ becomes important when  
that scale approaches the Kolmogorov dissipative length $l$. In this limit 
the really observed steepness of the energy spectra should significantly 
exceed the Kolmogorov value of $-5/3$. In the opposite case, if the observation 
scale $A$ approaches the system size $L$, the steepness also increases for a 
large scale aperture can hardly resolve the energy-containing range fluctuations. 
If the observation scale belongs to the inertial range, the steepness is utmost equal to the Kolmogorov value, and does not significantly depend on scale \cite{AHHS1998}. 

\section*{Acknowledgments} The authors are thankful to Drs O.Chkhetiani and E.Golbraikh, and to anonymous referees for useful comments, and to V.A.Krylov for his help with programming. The authors are also indebted to John Hopkins Turbulence Database Team for their kind permission to use the data of their simulations.  

The work was supported in part by the project "Monitoring"  of Space Research Institute RAS, by the Ministry of Education of Slovak Republic via grant VEGA 1/0345/17, and by the Ministry of Education and Science of Russian Federation  in the framework of Increase Competitiveness Program of MISIS.
%

\appendix
\section{Large scale stirring force \label{asf:sec}}
According to the Kolmogorov hypotheses, the energy 
injection should be equal to viscous energy dissipation per unit of mass. For stationary turbulence, to justify that the rate of energy 
dissipation per unit of mass
\begin{equation}
\varepsilon = \frac{d}{dt} \frac{\bra \vu^2(t,\vx)\ket}{2}  = \bra \dot{u}_i(t,\vx)u_i(t,\vx)\ket
\end{equation}
is compensated by the stirring force $f$, we chose the random force 
correlator by defining the correlation function of its scale components:
\begin{widetext}
$$
\langle \tilde{f}_{ai}(t,\vk) \tilde{f}_{a'j}(t',\vk') \rangle =
\delta(t-t') P_{ij} (\vk) g_0 \nu_0^3 C_g (2\pi)^d   
\delta^d(\vk+\vk') a \delta (a-a') \delta (a-L),  
$$
\end{widetext}
where $g_0$ is formal dimensionless strength of forcing, to be used for the 
perturbation expansion \citep{AAV1996}. The delta-functions in scale arguments ensure that 
fluctuations of different scales $a$ and $a'$ are uncorrelated, and the work 
is exerted  over the fluid only at large scale $a=L$. 

Reconstruction of Eulerian velocities $u_i$ from their scale components $u_{ai}$ by means of inverse wavelet transform gives
\begin{widetext}
\begin{align}\nonumber 
\bra \dot{u}_i(t,\vx)u_i(t,\vx)\ket &=& \frac{1}{C_g^2} \int \exp(\imath\vx(\vk_1+\vk_2)) \tilde{g}(a_1\vk_1) \tilde{g}(a_2\vk_2)
\bra \tilde f_{a_1i}(t,\vk_1) \int^t \tilde f_{a_2i}(\tau,\vk_2)d\tau \ket 
\frac{d^d\vk_1}{(2\pi)^d}\frac{da_1}{a_1} 
\frac{d^d\vk_2}{(2\pi)^d}\frac{da_2}{a_2} \\
&=& \frac{g_0\nu_0^3}{L^{d+1}} \frac{(d-1)}{2C_g} 
\int_0^\infty |\tilde{g}(\vy)|^2 \frac{d^d\vy}{(2\pi)^d}, \label{WW}
\end{align}
\end{widetext}
where factor $(d-1)$ comes from the trace of orthogonal projector, and the 
factor $\frac{1}{2}$ is the value of the $\theta$-function at the discontinuity.
 For particular case of $$\tilde{g}_1(k)=-\imath \vk e^{-\vk^2/2}$$ wavelet in $d=3$ dimensions, we get 
$$\quad C_g = \frac{1}{2}, \quad 
\int_0^\infty y^2 e^{-y^2} \frac{4\pi y^2 dy}{8\pi^3} = \frac{3}{16\pi^{3/2}}
\equiv \frac{\chi[g_1]}{2},
$$
Substituting this integral into \eqref{WW} one gets 
\begin{equation}
\varepsilon = \frac{g_0\nu_0^3}{L^4} \frac{3}{8\pi^{3/2}}. 
\end{equation}
\section{Feynman diagram technique \label{afd:sec}}
Using the generating functional \eqref{gfaa} with the action \eqref{act2} we can easily derive the Feynman diagram technique for the scale-dependent fields $\Phi_a$. 

The correlation functions are given by functional derivatives 
$$
\langle \Phi_{a_1}(x_1)\ldots \Phi_{a_n}(x_n)\rangle_c = \left.
\frac{\delta^n W[A]}{\delta A_{a_1}(x_1)\ldots \delta A_{a_n}(x_n)}
\right|_{A=0}.
$$
The difference is that each spatial integration measure $dx$ is substituted by the integration measure over affine group $\frac{dxda}{a}$, with functional  derivatives taken with respect to this measure.

Using the Fourier transform 
$$
u(x) = \int e^{\imath(\vk\vx -\omega t)}u(k) \frac{d^d\vk d\omega}{(2\pi)^{d+1}}
$$
of the fields in \eqref{act2} we make each convolution with basic wavelet $g$ into a multiplicative factor $\tilde{g}(a\vk)$. In this way we obtain the following diagram technique:
\begin{itemize}
\item Each external line is labeled by a pair $(a,k)$ (scale,momentum), and a vector index ($i$).
\item The integration in each internal line is performed over the measure $\frac{d\omega}{2\pi} \frac{d^d\vk}{(2\pi)^d} \frac{da}{a} \frac{1}{C_g}$.
\item There are two type of lines: a) Green functions 
$\bra uu' \ket$ and b) correlation functions $\bra uu\ket$; The auxiliary field $u'$ has zero moments $\bra u'u'\ket=0$. These Green functions are given by propagator matrix $K^{-1}$ multiplied by wavelet factors $\tilde{g}(a\vk)$ on each leg.
\item Each line carrying momentum $k$ is proportional to orthogonal projector 
$P_{ij}(\vk)$, where $i$ and $j$ are vector indices of the line, i.e. 
$$
G^{(0)}_{i\alpha,j\beta}(k) = \frac{\tilde{g}(\alpha\vk)P_{ij}(\vk) \overline{\tilde{g}}(\beta\vk)}{-\imath \omega + \nu_0 \vk^2} 
$$
for the Green function, and 
$$
D^{(0)}_{i\alpha,j\beta}(k) = P_{ij}(\vk) \frac{g_0 \nu_0^3}{C_g L}
\frac{\tilde{g}(\alpha\mathbf{k}) |\tilde{g}(kL)|^2\bar{\tilde{g}}(\beta\mathbf{k})}{|-\imath \omega + \nu_0 \vk^2|^2}  
$$
for the bare velocity pair correlation function.
\item Each vertex of the diagram is given by 
 $m_{abc}(k) = \frac{\imath}{2}\left(k_b \delta_{ac} + k_c \delta_{ab}\right),$
 multiplied by 3 wavelet factors of adjusted lines.
 
\item We consider the observation scale $A$ to be much bigger than the 
viscous dissipation scale $l$ and to belong the Kolmogorov range: $\xi = A/L \ll 1$.
For this reason we assume the statistical momenta of the turbulent velocity 
field are determined by direct energy cascade. {\em At the language of Feynman 
diagrams this means it should be no scales $a_i$ in internal lines less than 
minimal scale of all external lines.} 
\end{itemize}

\section{One-loop contributions to the Green functions \label{agf2:sec}}
The scale dependence of the viscosity is given by renormalization of the Green function $\bra \Phi_\alpha(x) \Phi_{\beta}(x') \ket$ by means of loop corrections.
In Fourier space the value of the Green function $G^{(2)}_{\alpha a,\beta s}(k)$, shown in diagram Fig.~\ref{gf1l:pic}, can be evaluated from 
the above mentioned expressions for the vertices and the Green functions 
after the integration over internal line scale arguments. The upper line 
of the diagram Fig.~\ref{gf1l:pic}, ($p^+$), contains the random stirring force correlator, the bottom line ($p^-$) is the Green function. 

The frequency integration in the loop integral 
can be done explicitly
\begin{align*}
\int_{-\infty}^\infty \frac{dq_0}{2\pi} 
|G_0(\frac{k}{2}+q)|^2 G_0(\frac{k}{2}-q) &=& 
\frac{1}{2\nu_0^2(\frac{\vk}{2}+\vq)^2} \times \\
&\times& \frac{1}{-\frac{\imath k_0}{\nu_0^2}+\frac{\vk^2}{2}+2\vq^2}
\end{align*}
where $$
G_0(k) = \frac{1}{-\imath k_0 + \nu_0 \vk^2}.$$

The loop tensor structure $L_{as}(k,p^+,p^-)$ \eqref{ts1} is given by 
the convolution of  
\begin{equation}
L(k,p^+,p^-) = m_{abc}(k) m_{fls}(p^-) P_{bl}(p^+)P_{cf}(p^-), 
\end{equation}
where 
$$ m_{abc}(k) = \frac{\imath}{2}\left(k_b \delta_{ac} + k_c \delta_{ab} \right),
\quad  
p^\pm = \frac{k}{2} \pm q,
$$
over repeated indices. 
The explicit equation for tensor structure  is  equation \eqref{ts1}. 

Let $A = \min(\alpha,\beta)$ be the minimal scale of two external 
lines of the diagram Fig.~\ref{gf1l:pic}. The lower line ($p^-$) contributes two identical $g_1$ wavelet factors 
$$
e^{-(A\vp^-)^2} = \frac{1}{C_g}\int_A^\infty |\tilde{g}(a\vp^-)|^2\frac{da}{a},
$$
so, the factor $e^{-2(A\vp^-)^2}$ will be prescribed to the bottom line; and similar factor $e^{-2(A\vp^+)^2}$ to the upper line.

Introducing the dimensionless momentum $\vq = |\vk|\vy$,  
with the polar angle between $\vq$ and $\vk$ measured from the $\vk$ direction, we can write the whole one-loop integral 
in $d=3$ dimension in the form 
\begin{widetext}
\begin{align*}\nonumber 
-\Sigma_{as}(k) &=& \frac{g_0\nu_0^3}{C_g L}\int k^3 \frac{y^2 dy}{(2\pi)^3} 
\sin\theta d\theta d\varphi L_{as}(k,p^+,p^-) 
\frac{1}{\nu_0^2 k^4} \frac{1}{\frac{1}{2}+2y^2+2y\cos\theta}
\frac{1}{\frac{1}{2}+2y^2 - \frac{\imath k_0}{\nu_0 k^2}}
\times \\
&\times& 
(L k)^2 \left[\frac{1}{4} + y^2 + y \cos \theta \right]
e^{-(L k)^2 \left[\frac{1}{4} + y^2 + y \cos \theta \right]}
e^{-2 (Lk)^2 \xi^2 \left[\frac{1}{4} + y^2 - y \cos \theta \right]}
e^{-2 (Lk)^2 \xi^2 \left[\frac{1}{4} + y^2 + y \cos \theta \right]}.
\end{align*}
\end{widetext}
This can be simplified to 
\begin{widetext}
$$
-\Sigma_{as}(k)= \frac{g_0\nu_0^3 C_g^{-1} Lk}{4\nu_0^2} \int \frac{y^2 dy}{(2\pi)^3} 
\sin\theta d\theta d\varphi L_{as}(k,p_+,p_-)
\frac{1}{\frac{1}{4}+y^2 - \frac{\imath k_0}{2\nu k^2}}
$$
\end{widetext}
The sign $-\Sigma$ is introduced for the value of one-loop integral is equal 
to minus ''self-energy'' contribution.

In view of the existence of only one preferred direction, that is the direction of $\mathbf{k}$, the integrals of $L_{as}$ should depend upon two scalars only: 
\begin{equation}
T_1 \equiv {\rm Tr}L_{as}, \quad T_2 = \frac{k_a k_s}{k^2} L_{as}
\end{equation}
\begin{align*}
T_1  &=& -\frac{(p^-k)(p^+p^-)^2}{2(p^+)^2 (p^-)^2} 
+ \frac{(p^+ k)(p^+ p^-)}{(p^+)^2} - \frac{(p^-k)}{2}, \\
T_2 &=& - \frac{(p^+k)(p^+p^-)(p^-k)^2}{(p^+)^2(p^-)^2 k^2}
+ \frac{(p^-k)(p^+k)^2}{2(p^+)^2 k^2} \\
&+& \frac{(p^+k)(p^+p^-)k^2}{2(p^+)^2 k^2}
+ \frac{(p^-k)^3}{(p^-)^2 k^2} - (p^-k).
\end{align*}
Assuming 
\begin{equation}
\Sigma_{as} = \nu_0 g_0 \Sigma^\delta \delta_{as} k^2 + \nu_0 g_0 \Sigma^T k_a k_s = -\int L_{as} d\mu(y,\theta),
\end{equation}
we get 
\begin{align*}
(d \Sigma^\delta + \Sigma^T) k^2 &=& -\int T_1 d\mu(y,\theta) = -k^2 I_1, \\
(\Sigma^\delta + \Sigma^T) k^2 &=& -\int T_2 d\mu(y,\theta) = -k^2 I_2, 
\end{align*}
where $d\!=\!3$ is the space dimension. So, we need to calculate $I_1$ and $I_2$. 
From where 
$$
\Sigma^{\delta} = -\frac{I_1-I_2}{2}, \quad \Sigma^T = -I_2 - \Sigma^\delta=-\frac{3I_2-I_1}{2}.$$
The substitution $\mathbf{q} = k \mathbf{y}, (kq) = ky\mu$, with $\mu \equiv \cos\theta$, gives:
\begin{widetext}
\begin{align*}
T_1 &=& {\rm Tr} L_{as} = k^2 \frac{-\mu^3 y^3 + 2\mu^2 y^4 + \mu y^3 - 2y^4}{2\left[\left(\frac{1}{4}+y^2 \right)^2 - y^2\mu^2 \right]}
=\frac{8k^2(\mu-1)(\mu+1)(2y-\mu)y^3}{(4y^2-4y\mu+1) (4y^2+4y\mu+1)},\\
T_2 &=& \frac{k_ak_s}{k^2}L_{as} = k^2 \frac{-\mu^3 y^5+\mu y^5 -\mu^4 y^4 +\frac{5}{2}\mu^2y^4-\frac{3}{2}y^4+\frac{1}{4}\mu^3 y^3 - \frac{1}{4}\mu y^3-\frac{1}{8}\mu^2y^2+\frac{1}{8}y^2}{2(1/4+y\mu+y^2) (1/4-y\mu+y^2)} \\
&=& -k^2 \frac{(\mu-1)(\mu+1)(8\mu y^3+8\mu^2y^2-12y^2-2\mu y +1)}{ (4y^2-4y\mu+1) (4y^2+4y\mu+1)} y^2,\\
\frac{T_1-T_2}{2} &=& k^2 \frac{(\mu^2-1) y^2 (8\mu^2y^2 + \mu (8y^3-10y)  + 4y^2 +1)}{2 (4y^2-4y\mu+1) (4y^2+4y\mu+1)}, \\
\frac{3T_2-T_1}{2} &=& -k^2 \frac{(\mu^2-1) y^2 (24\mu^2y^2 + \mu (24y^3-14y)  - 20y^2 +3)}{2 (4y^2-4y\mu+1) (4y^2+4y\mu+1)}.
\end{align*}
\end{widetext}
After standard algebraic manipulations, this gives: 
\begin{widetext}
\begin{align}
\Sigma^\delta &=& \frac{kL}{128 C_g} \int_0^\infty \frac{y^2 dy}{(2\pi)^2}
\frac{e^{-(kL)^2(1+4\xi^2) \left(\frac{1}{4}+y^2 \right)}}{\frac{1}{4}+y^2-\imath\frac{2k_0}{\nu_0\vk^2}} \int_{-1}^{1}d\mu
\frac{(1-\mu^2)(8\mu^2y^2+\mu(8y^3-10y)+4y^2+1)}{\left(\frac{1}{4y}+y-\mu \right)\left(\frac{1}{4y}+y+\mu \right) } e^{-(kL)^2 y\mu}, \label{Sigma-D}\\
\nonumber \Sigma^T &=& -\frac{kL}{128C_g} \int_0^\infty \frac{y^2 dy}{(2\pi)^2}
\frac{e^{-(kL)^2(1+4\xi^2) \left(\frac{1}{4}+y^2 \right)}}{\frac{1}{4}+y^2-\imath\frac{2k_0}{\nu_0\vk^2}} \int_{-1}^{1}d\mu
\frac{(1-\mu^2)(24\mu^2y^2+\mu(24y^3-14y)-20y^2+3)}{\left(\frac{1}{4y}+y-\mu \right)\left(\frac{1}{4y}+y+\mu \right) } e^{-(kL)^2 y\mu}, 
\end{align} 
\end{widetext}
where $\mu=\cos\theta$ determines the polar angle between $\vk$ and $\vq$.
Since the velocity field is transversal, it is natural to present the 
self-energy as a sum of transversal and longtitudal terms:
\begin{equation}
\Sigma_{as} = \nu g \Sigma^{\delta} \left(\delta_{as}-\frac{k_ak_s}{\vk^2} \right)\vk^2 + \nu g\Sigma^{l} k_a k_s,
\end{equation} 
where $\Sigma^{l}=\Sigma^\delta+\Sigma^T$.

\section{One-loop contribution to stirring force correlator\label{adf:sec}}
The one-loop contribution to stirring force correlator is shown in Fig.~\ref{gf2l:pic} below. 
\begin{figure}[ht]
\centering \includegraphics[width=4cm]{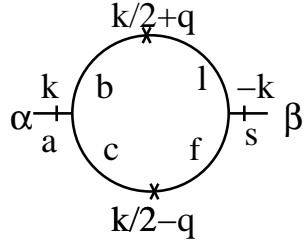}
\caption{One-loop contribution to the correlation function. Greek letters $\alpha$ and $\beta$ denote scale arguments.}
\label{gf2l:pic}
\end{figure}
The tensor structure of this diagram has the form 
\begin{equation}
C_{as}(k,p^+,p^-) = -m_{abc}(k) m_{sfl}(k) P^{bl}(\vp^+)P^{cf}(\vp^-), \label{Cpp}
\end{equation}
or explicitly:
\begin{align*}
C_{as}(k,p^+,p^-) &=& \delta_{as} \left[ k^2 - \frac{(p^+k)^2}{4(p^+)^2} - \frac{(p^-k)^2}{4(p^-)^2} \right] 
+ \frac{k_ak_s}{2} \\ 
 &+& \frac{p^+_a p^+_s}{4} \left[  \frac{(p^-k)^2}{(p^-)^2(p^+)^2} - \frac{k^2}{(p^+)^2} \right] \\
&+& \frac{p^-_a p^-_s}{4} \left[  \frac{(p^+k)^2}{(p^-)^2(p^+)^2} - \frac{k^2}{(p^-)^2} \right] \\
&=&  (p^+_a p^-_s + p^-_a p^+_s) \frac{(p^+k)(p^-k)}{4(p^+)^2(p^-)^2} \\ 
&-& (k_ap^+_s + p^+_a k_s) \frac{(p^+k)}{4(p^+)^2}\\
&-& (k_ap^-s + p^-_a k_s) \frac{(p^-k)}{4(p^-)^2}. 
\end{align*}
The corresponding invariants, the tensor structure can depend on, are 
\begin{align*}
C_1 &=& {\rm Tr}C_{as} &=& \frac{3}{2}k^2 - \frac{(p^-k)^2}{(p^-)^2} 
- \frac{(p^+k)^2}{(p^+)^2} \\
 & &  &+& \frac{(p^+k)(p^-k)(p^+p^-)}{2(p^+)^2(p^-)^2}, \\
C_2 &=& \frac{k^ak^s}{k^2}C_{as} &=& k^2 -\frac{(p^-k)^2}{(p^-)^2} 
-\frac{(p^+k)^2}{(p^+)^2} \\
& & &+& \frac{(p^-k)^2(p^+k)^2}{(p^-)^2(p^+)^2}.
\end{align*} 
In dimensionless variables
\begin{align}\nonumber 
C_1 &=& -\frac{2k^2 (\mu^2-1)y^2(12y^2+1)}{(4y^2-4y\mu +1) (4y^2+4y\mu +1)}, \\
\nonumber C_2 &=& \frac{16k^2 (\mu^2-1)^2 y^4}{(4y^2-4y\mu +1) (4y^2+4y\mu +1)}, \\
\nonumber \frac{C_1-C_2}{2} &=& -\frac{k^2 (\mu^2-1)y^2(8y^2\mu^2+4y^2+1)}{(4y^2-4y\mu +1) (4y^2+4y\mu +1)},\\
\frac{3C_2-C_1}{2} &=& \frac{k^2 (\mu^2-1)y^2(24y^2\mu^2-12y^2+1)}{(4y^2-4y\mu +1) (4y^2+4y\mu +1)}. \label{cdelta}
\end{align}

The total symmetric contribution to the 1PI correlation function has the 
form 
\begin{align*}
\Delta^{(2)}_{as} &=& \left(\frac{g_0\nu_0^3}{L C_g}\right)^2 \int 
\frac{d^4 q}{(2\pi)^4}
\frac{|\tilde{g}(Lp^+)|^2 f^2(Ap^+)}{|-\imath\omega^+ + \nu_0 (p^+)^2|^2} \\
&\times& \frac{|\tilde{g}(Lp^-)|^2 f^2(Ap^-)}{|-\imath\omega^- + \nu_0 (p^-)^2|^2} 
  C_{as}(k,p^+,p^-).
\end{align*} 
In case the non-zero frequency  $k_0 \ne 0$
\begin{widetext}
\begin{equation}
\int_{-\infty}^\infty \frac{dq_0}{2\pi} 
\left| G_0\left(\frac{k}{2}+q \right)\right|^2
\left| G_0\left(\frac{k}{2}-q \right)\right|^2 = 
\frac{1}{4\nu_0^3} 
\frac{
\frac{\mathbf{k}^2}{4}+\mathbf{q}^2
}
{
\left(
\frac{\mathbf{k}^2}{4}+\mathbf{q}^2 + \imath \frac{2k_0}{\nu_0}
\right)
\left(
\frac{\mathbf{k}^2}{4}+\mathbf{q}^2 - \imath \frac{2k_0}{\nu_0}
\right)
}
\cdot
\frac{1}{
\frac{\mathbf{k}^2}{4} +\mathbf{q}^2 + \mathbf{k}\mathbf{q}
} \cdot 
\frac{1}{
\frac{\mathbf{k}^2}{4} +\mathbf{q}^2 - \mathbf{k}\mathbf{q}
}.
\end{equation}
\end{widetext}
In case of the zero frequency $k_0=0$ this turns to be 
\begin{widetext} 
\begin{equation}
\int_{-\infty}^\infty \frac{dq_0}{2\pi} 
\left| G_0\left(\frac{k}{2}+q \right)\right|^2
\left| G_0\left(\frac{k}{2}-q \right)\right|^2 = 
\frac{1}{4\nu_0^3} \frac{1}{\frac{\mathbf{k}^2}{4}+\mathbf{q}^2}
\cdot
\frac{1}{
\frac{\mathbf{k}^2}{4} +\mathbf{q}^2 + \mathbf{k}\mathbf{q}
} \cdot 
\frac{1}{
\frac{\mathbf{k}^2}{4} +\mathbf{q}^2 - \mathbf{k}\mathbf{q}
}.
\end{equation}
\end{widetext}
The $\delta$-function in scale variables results in square of wavelets 
in both upper and lower parts of the loop
\begin{align*}
|\tilde{g}(Lp^+)|^2 |\tilde{g}(Lp^-)|^2 &=& L^2 \left(\frac{\mathbf{k}}{2}+\mathbf{q} \right)^2
e^{
-L^2 \left(\frac{\mathbf{k}}{2}+\mathbf{q} \right)^2
}\\
&\times& 
L^2 \left(\frac{\mathbf{k}}{2}-\mathbf{q} \right)^2
e^{
-L^2 \left(\frac{\mathbf{k}}{2}-\mathbf{q} \right)^2
}
\end{align*}
The angle arguments in the exponents are canceled due to symmetry.
The filter factors for $g_1$ wavelet are 
$$f(Ap^\pm) = e^{-(Ak)^2 (1/4 + y^2 \pm y \cos \theta)}$$  
After integration over the loop frequency variable $\int_{-\infty}^\infty \frac{dq_0}{2\pi} \ldots $ we get 
\begin{widetext}
\begin{align*}
\Delta^{(2)}_{as} = \left(\frac{g_0\nu_0^3}{L C_g}\right)^2 
(kL)^4 \int 
\left[ \left(\frac{1}{4} + y^2 \right)^2 -y^2 \cos^2\theta 
\right]
e^{-2(kL)^2 \left(\frac{1}{4}+y^2\right)} 
\frac{1}{4\nu_0^3}
\frac{
\frac{\mathbf{k}^2}{4}+\mathbf{q}^2
}
{
\left(
\frac{\mathbf{k}^2}{4}+\mathbf{q}^2 + \imath \frac{2k_0}{\nu_0}
\right)
\left(
\frac{\mathbf{k}^2}{4}+\mathbf{q}^2 - \imath \frac{2k_0}{\nu_0}
\right)
}
\cdot \\
\frac{1}{
\frac{\mathbf{k}^2}{4} +\mathbf{q}^2 + \mathbf{k}\mathbf{q}
} \cdot 
\frac{1}{
\frac{\mathbf{k}^2}{4} +\mathbf{q}^2 - \mathbf{k}\mathbf{q}} 
e^{-4\xi^2(Lk)^2 \left(\frac{1}{4} + y^2 \right)} C_{as}(k,p^+,p^-)
\frac{q^2 dq}{(2\pi)^2} \sin\theta d\theta. 
\end{align*}
\end{widetext}
After algebraic simplification we get  
\begin{widetext}
\begin{align*}
\Delta^{(2)}_{as}(L,k) &=& \left( \frac{g_0\nu_0^3}{2C_g}
\right)^2 \frac{L^2}{\nu_0^3}\int 
\frac{q^2 dq}{(2\pi)^2}\sin\theta d\theta 
\frac{e^{-2 (Lk)^2 (1+2\xi^2) \left(\frac{1}{4}+y^2\right)}C_{as}(k,p^+,p^-)
\left( \frac{\mathbf{k}^2}{4}+\mathbf{q}^2 \right)
}
{
\left(
\frac{\mathbf{k}^2}{4}+\mathbf{q}^2 + \imath \frac{2k_0}{\nu_0}
\right)
\left(
\frac{\mathbf{k}^2}{4}+\mathbf{q}^2 - \imath \frac{2k_0}{\nu_0}
\right)
}.
\end{align*}
\end{widetext}
In view of isotropy, using equation \eqref{cdelta}, we get 
$$
C_{as} = C^{\delta} \delta_{as} + C^T \frac{k_a k_s}{k^2}
$$
with
\begin{widetext}
\begin{align}
C^{\delta}(L,k) &=& \left( \frac{g_0\nu_0^3}{2C_g}
\right)^2 \frac{L^2 k^3}{16\nu_0^3}\int 
\frac{y^4 dy}{(2\pi)^2}d\mu  
\frac{e^{-2 (Lk)^2 (1+2\xi^2) \left(\frac{1}{4}+y^2\right)} 
\left( \frac{1}{4}+y^2 \right)
}
{
\left(
\frac{1}{4}+y^2 + \imath \frac{2k_0}{\nu_0 \mathbf{k}^2}
\right)
\left(
\frac{1}{4}+y^2 - \imath \frac{2k_0}{\nu_0\mathbf{k}^2}
\right)
}
\frac{(1-\mu^2)(8y^2\mu^2+4y^2+1)}{
\left(
\frac{1}{4}+y^2 -y\mu
\right)
\left(
\frac{1}{4}+y^2 +y\mu
\right)
}.
\end{align}
\end{widetext}
\section{Renormalization group equations \label{arge:sec}}
\subsection{Renormalization of viscosity \label{anu:sec}}
We use the following set of scales 
$
l=A_0< A_1< A_2< \ldots< A_L=L
$.
In terms of dimensionless variable $\xi=A/L$ this corresponds to
$
\xi_0=R^{-1},\xi_1,\xi_2,\ldots, \xi_L=1,$ where $R=L/l \gg 1$.
If we use the iteration from the Kolmogorov dissipation scale $l$ to 
the macro scales, we would get the following chain of equations: 
\begin{align*}
\nu_1  &=& \nu_0 [1 + g_0 \Sigma(\xi_0)], \\
\nu_2  &=& \nu_1 [1 + g_1 \Sigma(\xi_1)], \\
\nu_{k+1} &=& \nu_k [1 + g_k \Sigma(\xi_{k})].
\end{align*}
If we go from large scales to smaller scales the proposed 
inversion formulae are: 
\begin{align}
\nonumber \nu_0  &\approx& \nu_1 [1 - g_1 \Sigma(\xi_0)], \\
\nonumber \nu_1  &\approx& \nu_2 [1 - g_2 \Sigma(\xi_1)], \\
\nu_{k-1} &\approx& \nu_k [1 - g_k \Sigma(\xi_{k-1})], \label{itt2}
\end{align}
where $g_k \equiv g(\xi_k)$.
The iteration scheme above can be written in a form of difference equation 
\begin{equation}
\frac{\nu_{k-1}-\nu_k}{\nu_k}= -g(\xi_k) \Sigma(\xi_{k-1}).
\end{equation}
For the equal scale steps 
$
A_k = A_0 \delta^k, \xi_k = \xi_0 \delta^k, \Delta \ln \xi = \ln\delta \Delta k,$ we get 
$$
\frac{\Delta \ln \nu}{\Delta k} = g(\xi_k) \Sigma(\xi_k),\quad \hbox{or\ }
\frac{d\ln\nu}{d\ln\xi}=g(\xi) \frac{\Sigma(\xi)}{\ln\delta}.
\eqno{(\ref{rge-nu})}
$$
\subsection{Renormalization of stirring force}
The one-loop contribution to the stirring force correlation function is shown in Fig.~\ref{gf2l:pic}. 
 
Since the stirring force acts on large 
scales only, its yield on smaller scales will be the sum of correlator 
itself and the one loop correction 
\begin{equation}
D_{L-1} = D_L + D_L^2 * OneLoopK(\xi_{L-1})
\end{equation}
with $D(\xi) = g(\xi)\nu^3(\xi)/L$, hence  
$$
\frac{D_{L-1}-D_L}{D_L} = D_L * OneLoopK(\xi_{L-1}).$$
In differential form the latter difference equation yields 
$$
\frac{d\ln D}{d\ln\xi} = - \frac{K(\xi)}{\ln\delta} 
\eqno{(\ref{rge-D})},
$$
where 
\begin{widetext}
\begin{align}
K(\xi) &=& \frac{(kL)^3}{16}\int 
\frac{y^4 dy}{(2\pi)^2}d\mu  
\frac{e^{-2 (Lk)^2 (1+2\xi^2) \left(\frac{1}{4}+y^2\right)} 
\left( \frac{1}{4}+y^2 \right)
}
{
\left(
\frac{1}{4}+y^2 + \imath \frac{2k_0}{\nu_0 \mathbf{k}^2}
\right)
\left(
\frac{1}{4}+y^2 - \imath \frac{2k_0}{\nu_0\mathbf{k}^2}
\right)
}
\frac{(1-\mu^2)(8y^2\mu^2+4y^2+1)}{
\left(
\frac{1}{4}+y^2 -y\mu
\right)
\left(
\frac{1}{4}+y^2 +y\mu
\right)
}.
\end{align}
\end{widetext}
The value of $K(\xi)$ is typically a few orders of magnitude 
less than $\Sigma(\xi)$. 
\section{Frequency integrals that contribute to velocity pair correlator \label{uu:sec}}
The bare correlation function, being integrated over the frequency, contains 
the integral
\begin{equation}
I_{w0} = \int_{-\infty}^\infty \frac{d\omega}{2\pi} \frac{1}{|-\imath\omega+A|^2} = \frac{1}{2A}, \quad A = \nu \mathbf{k}^2
\end{equation}
The integral coming from 1PI correlation function multiplied by 
two conjugated correlation functions from the legs of the diagram has the form
\begin{equation}
I_{w2} = \int_{-\infty}^\infty \frac{d\omega}{2\pi} \frac{1}{|-\imath\omega+A|^2} \frac{1}{|B+\imath\frac{2\omega}{A}|^2}, 
\quad B = \frac{1}{4} + y^2.
\end{equation} 
Substituting $C = \frac{AB}{2}$, we get 
\begin{align*}
I_{w2} &=& \int_{-\infty}^\infty \frac{d\omega}{2\pi} \frac{A^2}{4}
\frac{1}{\omega^2 + A^2} \frac{1}{\omega^2 + C^2} = \frac{1}{8} \frac{A}{C(A+C)} \\ &=& \frac{1}{4 \left(\frac{1}{4}+y^2 \right)\nu\mathbf{k}^2 \left( 1+\frac{1}{2}\left(\frac{1}{4}+y^2 \right)\right)}.
\end{align*}
The third integral comes from two conjugated diagrams with self-energy multiplied by correlator and propagator 
\begin{align*}
I_{1w}  &=& \int_{-\infty}^\infty \frac{d\omega}{2\pi}
\frac{1}{\omega^2+A^2} \cdot \frac{1}{-\imath\omega+A}\cdot	\frac{2A}{4C-\imath\omega} \\ &=& \frac{1}{2A(A+4C)} = \frac{1}{2(\nu\mathbf{k}^2)^2 \left(
1+2 \left(\frac{1}{4}+y^2 \right)\right)}.
\end{align*}
This integral contributes twice for there are two conjugated diagrams, shown 
in the lower part of Fig.~\ref{c2:pic}. 
\end{document}